\documentclass[a4paper,11pt]{article}
\usepackage{pos}

\title{Data-driven extrapolation schemes of {\em Fermi}-LAT spectra to the VHE}
 \ShortTitle{Fermi-LAT extrapolations to VHE}

\author*[a,b,c]{M. Nievas Rosillo}
\author[c,d]{T. Hassan Collado}

\affiliation[a]{Instituto de Astrof\'isica de Canarias (IAC) \\ C/Via Lactea S/N E-38205 La Laguna, Tenerife, Spain}

\affiliation[b]{Universidad de La Laguna, Dept. Astrof\'isica (ULL) \\
Av. Astrofisico Francisco Sánchez, S/N, E-38206 La Laguna, Tenerife, Spain}

\affiliation[c]{Deutsches Elektronen-Synchrotron (DESY) \\ Platanenallee 6, Zeuthen, Germany, Spain}

\affiliation[d]{Centro de Investigaciones Energéticas, Medioambientales y Tecnológicas (CIEMAT)\\
Av. Complutense, 40, E-28040 Madrid, Spain}

\emailAdd{mnievas@iac.es}
\emailAdd{tarek.hassan@ciemat.es}

\abstract{After 10 years of operations of the Large Area Telescope (LAT), a high-energy pair-creation telescope onboard the Fermi satellite, the Fermi Collaboration has produced two major catalogs: the 4FGL and the 3FHL. These catalogs represent the best sample of potential very high energy (VHE) emitters that may be studied by Imaging Atmospheric Cherenkov Telescopes (IACTs). Several methods are used to extrapolate the Fermi-LAT spectra to TeV energies, generally using simple analytical functions. The recent success of IACTs has motivated the creation of catalogs listing the discoveries of these experiments. Among these initiatives, gamma-cat excels as an open-access tool to archive high-level results in the VHE field, such as catalogs, spectra and light curves. By using these resources, we present a data-driven methodology to test the reliability of different VHE extrapolation schemes used in the literature and evaluate their accuracy reproducing real VHE observations.}

\FullConference{37$^{\rm{th}}$ International Cosmic Ray Conference (ICRC 2021)\\
		July 12th -- 23rd, 2021\\
		Online -- Berlin, Germany}

\begin{document}
\maketitle

\section{Introduction}

The extragalactic $\gamma$-ray sky is dominated by the emission from Active Galactic Nuclei (AGN). Among AGN, of special relevance are blazars, a type of radio-loud sources which emit mostly non-thermal radiation and whose jets are closely aligned with the line of sight. Blazars can be subclassified \cite{1995PASP..107..803U} according to the presence or absence of optical spectral features in Flat Spectrum Radio Quasars (FSRQs, strong emission lines) and BL Lacertae objects (BL Lac, weak or absent emission lines). They can also be subdivided according to their Synchrotron emission peak frequency in \cite{2010ApJ...716...30A}: 

\begin{enumerate}
    \item Low Synchrotron Peaked (LSP): $\nu_{Sync} < 10^{14}\,\mathrm{Hz}$. 
    \item Intermediate Synchrotron Peaked (ISP): $10^{14} \nu_{Sync} < 10^{15}\,\mathrm{Hz}$.
    \item High Synchrotron Peaked (HSP): $10^{15} < \nu_{Sync} < 10^{17}\,\mathrm{Hz}$.
    \item Extreme High Synchrotron Peaked (EHSP): $\nu_{Sync}>10^{17}\,\mathrm{Hz}$.
\end{enumerate}

FSRQs are mostly IBL and LBLs, while BL Lacs belong to any of these four categories. Blazars (with the exception of, perhaps, EHSPs), are very variable sources across the entire electromagnetic spectrum.  

The emission of high energy ($E>100\,\mathrm{MeV}$) $\gamma$-rays from blazars has been studied successfully for decades, first with EGRET, onboard the Compton Gamma-Ray Observatory and later with LAT, onboard the {\em Fermi} Gamma-ray Space Telescope. Both have produced catalogs of $\gamma$-ray sources \citep{1999ApJS..123...79H, 2020ApJS..247...33A} covering the entire sky, together with specific AGN catalogs \cite{2020ApJ...892..105A}. While the sky exposure is not completely homogeneous, after several years of observations they provide the most unbiased source of information of the $\gamma$-ray extralactic sky available today.

Space telescopes are however not suited to observe beyond $\sim 100$ $\mathrm{GeV}$, the energy threshold usually used to define the Very High Energy (VHE) regime. Their limited collection area, of at most $\sim 1\,\mathrm{m^2}$, prevents these instruments to gather enough statistics over reasonable observation times. Instead, the VHE regime is best studied from Earth using arrays of Imaging Atmospheric Cherenkov Telescopes (IACT). The current generation of IACTs includes MAGIC and VERITAS in the Northern Hemisphere and H.E.S.S. in the Southern Hemisphere. The Cherenkov Telescope Array (CTA) represents the next generation of these instruments: it will dramatically improve both the sensitivity and resolution of the current generation of IACTs and provide full-sky coverage as it will be composed of two IACT arrays, one in each hemisphere. 

For the study of VHE $\gamma$-ray sources, IACTs pose a number of advantages. Their collection areas are $\sim 10^{5}\,\mathrm{m^2}$. This results in much better short-timescale sensitivity.
Stereoscopic operations also permit better angular and energy resolution. However, their limited field of view when compared with space detectors ($\lesssim 10^\circ$ diameter) often results in telescopes being externally triggered to observe blazar outbursts, biasing blazar observations towards flaring states. The assessment of how does the Extragalactic VHE Universe looks like becomes a complex problem, while being specially relevant to plan and predict the scientific prospects of the future CTA concerning VHE AGN populations. So far, two techniques have been explored:

\begin{itemize}
    \item Deep exposures on specific targets to determine their `quiescent' VHE emission \cite{2016ApJ...819..156B}. 
    \item Extrapolations from HE observations performed by {\em Fermi}-LAT \citep[e.g.][]{2021JCAP...02..048A}
\end{itemize}

Each have their own set of limitations: The first requires long and expensive observations on sources which may not even be detectable by the current generation of IACTs.
It is also needed to assume that the observed sources are representative of the entire sky, requiring extra effort to do an unbiased source selection. The second relies on our knowledge of how $\gamma$-ray emission is produced. If non-physical spectral models are used, the presence of possible spectral breaks or cut-offs is usually not accounted for. Finally, $\gamma$-ray opacity due to photon-photon interaction with the Extragalactic Background Light (EBL) needs to be estimated \cite{2008IJMPD..17.1515R}, requiring a good knowledge of the blazar distance (often unknown for BL Lacs).

This work suggests a hybrid approach to evaluate how accurately we can extrapolate HE spectra to VHEs based on real observations of blazars by IACTs.

\section{Method}


To perform our study in a systematic and reproducible way, we created a simple pipeline, named {\tt ext2TeV}, which stands for `extrapolation to TeV energies`. The source code is open and available at gitHub\footnote{\url{https://github.com/mireianievas/ext2TeV}} under a MIT license. It is written in Python 3.8 and uses standard modules such as numpy, scipy, astropy and matplotlib. 
The tool performs the following tasks: 

\begin{enumerate} 
\item Multiple catalog loading and source matching. 
\item Spectral point extraction from catalogs.
\item Spectral model evaluation using typical analytical shapes: Power Law, Log Parabola, Power Law with (sub-)exponential cut-off and Log Parabola with exponential cut-off.
\item Applied EBL absorption on spectral models.
\item Assessment of residuals for VHE data and extrapolated models, for different blazar spectral types and different spectral shapes.
\end{enumerate}

In the following sections we provide further details about these points.

\subsection{Catalogs}\label{sec:catalogs}

\subsubsection{High Energy regime}\label{sec:catalogs:he}

In the HE regime, {\tt ext2TeV} currently supports the 4FGL \cite{2020ApJS..247...33A} (results are reported for the DR2), 4LAC \cite{2020ApJ...892..105A} (DR2) and the 3FHL \cite{2017ApJS..232...18A} catalogs from {\it Fermi}-LAT. 4LAC is based on the same data as the 4FGL, hence has the same 8 years exposure (10 years for the DR2), but it focuses on extragalactic sources and adds additional properties for these sources such as redshift estimates or blazar type classifications. 4LAC contains 3207 sources (plus 285 additional ones added in 4LAC-DR2). The 3FHL on the other hand is dedicated to sources with hard spectra and significant emission above the catalog threshold of $10\,\mathrm{GeV}$, therefore representing an excellent sample of potential VHE emitters. Being it a more dated catalog, the total exposure is only 7 years. 1556 sources were reported in that catalog, being most of them ($\sim 79\%$) extragalactic sources. Both spectral points and fitted spectral models are extracted from 4FGL/4LAC, while only spectral points are used from the 3FHL.

\subsubsection{Very High Energy regime}\label{sec:catalogs:vhe}

For the VHE regime, we included two publicly-released catalogs: GammaCAT \cite{gammacat} and VTSCat \cite{vtscat}. GammaCAT defines itself as an open data collection and source catalog for gamma-ray astronomy and it is a {\tt gammapy} sub-project hosted in github: \url{https://github.com/gammapy/gamma-cat}. It aims at providing data taken by multiple experiments, including: HESS, MAGIC, VERITAS, WHIPPLE, HEGRA, CANGAROO, HAWC, MILAGRO, CRIMEA, ARGO, CAT and DURHAM. Unfortunately, the task of maintaining such a catalog is extremely demanding, so the catalog is incomplete (as of June 2021), with its last entries dating from 2018. VTSCat instead is a more recent development, focused on data collected and published by the VERITAS collaboration, mostly using the VERITAS telescopes. Its format resembles that of GammaCAT, including also a collection of spectra for many VHE sources. In total, we collected spectra for 43 AGN, most (25) being HSP blazars, 11 being ISP or LSP, two classified as EHSP and 5 of unknown type. For many of these AGN, there are multiple datasets published. 

Because most blazars are variable, in many cases we do not have one but several spectra for the same blazar, corresponding to different levels of VHE $\gamma$-ray activity,  different times or different instruments. In order to keep the study simple, we integrate the different spectra to seek for the minimum activity (lowest state), keeping more elevated states only for reference. The additional advantage of this is that it removes possible duplicity if the same dataset appears in both VTSCat and GammaCAT. To make the comparison of the lowest state in VHE to the average state recorded in HE possible, we assume that most sources spend most of the time in quiescent states and that elevated states are short and rather sporadic events with a small impact in the average flux.

\subsection{SED extrapolations}

The aim of this study is to test how reliably we can extrapolate HE spectra to the VHE regime. This is a necessary step in order to make predictions for existing and upcoming IACTs. Therefore, we use spectral shapes already fitted and included in the {\it Fermi}-LAT's catalogs of section \ref{sec:catalogs:he} (or shapes derived from them) and compare how accurately they reproduce the observed VHE SEDs from section \ref{sec:catalogs:vhe}. So far, we have implemented the following intrinsic spectral functions:

\begin{enumerate}
    \item Power Law (PWL): $\phi_{\rm int} (E_{\gamma}) = \phi_0 (E_\gamma/E_0)^{-\Gamma}$
    \item Log Parabola (LP): $\phi_{\rm int} (E_{\gamma}) = \phi_0 (E_\gamma/E_0)^{-\Gamma-\beta \ln (E_\gamma/E_0)}$
    \item PWL with (sub-)exponential cut-off (PLEC): $\phi_{\rm int} (E_{\gamma}) = \phi_0 (E_\gamma/E_0)^{-\Gamma} \exp(-(E_\gamma/E_{\rm cut})^{\alpha})$
    \item CTAGammaProp's model from  \citep{2021JCAP...02..048A}: $\phi_{\rm int} (E_{\gamma}) = \phi_0 (E_\gamma/E_0)^{-\Gamma-\beta \ln (E_\gamma/E_0)} \exp(-(E_\gamma/E_{\rm cut}))$
\end{enumerate}

The parameters of the first three models are directly extracted from the 4FGL catalog ($\alpha=2/3$ for most sources) while for the last model, we adopted 4FGL's Log Parabola model parameters and then set the cut-off energy to be $E_{\rm cut}= E^\prime_{\rm cut}/(1+z)$, where $z$ is the redshift of the source and $E^\prime_{\rm cut}$ is $100\,\mathrm{GeV}$ for ISP and LSPs, $1\,\mathrm{TeV}$ for HSPs and $10\,\mathrm{TeV}$ for EHSPs (recipe used in \citep{2021JCAP...02..048A}).

\subsection{EBL absorption}

VHE $\gamma$-rays can interact with background photons from the Extragalactic Background Light (EBL). This induces a `cosmic opacity` to VHE photons that depends on the energy of the $\gamma$-ray, the redshift of the emitter and the EBL intensity (that is predicted by a given EBL model). Since most of the emission observed by {\it Fermi}-LAT occurs at $<50\,\mathrm{GeV}$, it is basically unaffected by EBL absorption, at least for sources at redshifts $z\lesssim 1$. Therefore, the spectral models included in {\it Fermi}-LAT are predictors of the intrinsic, un-absorbed, emission from the source. In order to get predictors for the observed VHE emission, we need to include EBL absorption: $\phi_{\rm obs} (E_\gamma) = \phi_{\rm int} (E_\gamma) \exp(-\tau_{\rm EBL} (z,E_\gamma))$. Only the EBL model of \citet{2011MNRAS.410.2556D} is currently implemented within {\tt ext2TeV} at the moment.


\section{Results}

\begin{figure}
    \centering
    \includegraphics[width=.65\linewidth]{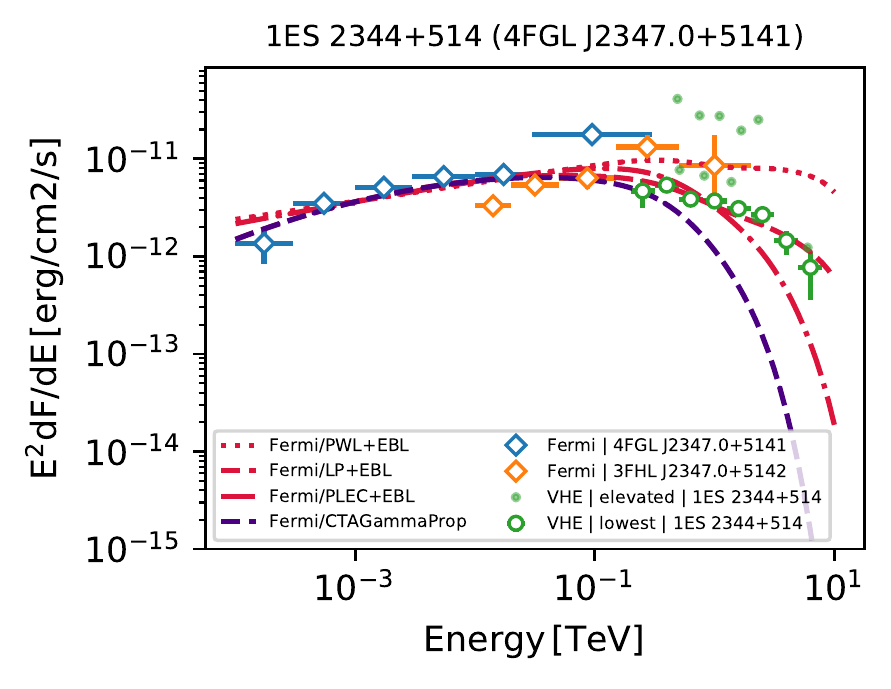}
    \includegraphics[width=.65\linewidth]{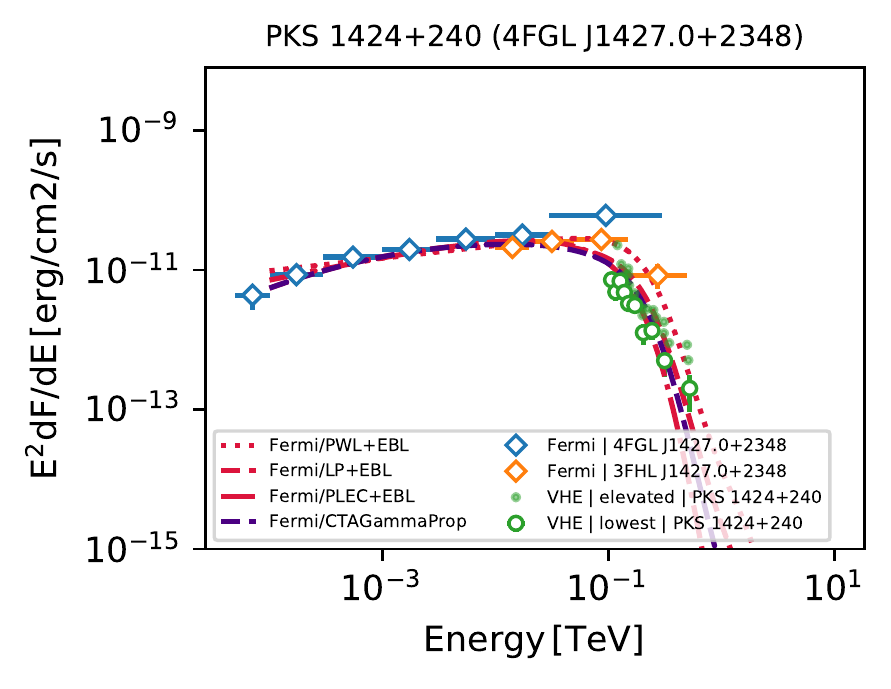}
    \caption{Spectral energy distribution collected from two {\it Fermi}-LAT catalogs plus three VHE spectra for the source 1ES~2344+514 ($z=0.044$) and PKS~1424+240. The different extrapolated models, only using spectral information from the HE regime, are shown in dashed/dotted lines.}
    \label{fig:spectrum}
\end{figure}

Figure \ref{fig:spectrum} shows, as an example, the HE and VHE spectra collected from a couple of the 43 blazars considered in the study. It contains {\it Fermi}-LAT spectral points as they appear in the 4FGL and 3FHL catalogs, together with spectra of different VHE datasets. The lowest state is identified (open green circles). In the first case (an HSP) it is best reproduced by a LP model plus EBL absorption, while the second (ISP) is reproduced accurately by most models excepts, perhaps, the PWL. The other three spectral models either overestimate or underestimate the flux emitted in the TeV band. 


\begin{figure}
    \centering
    \includegraphics{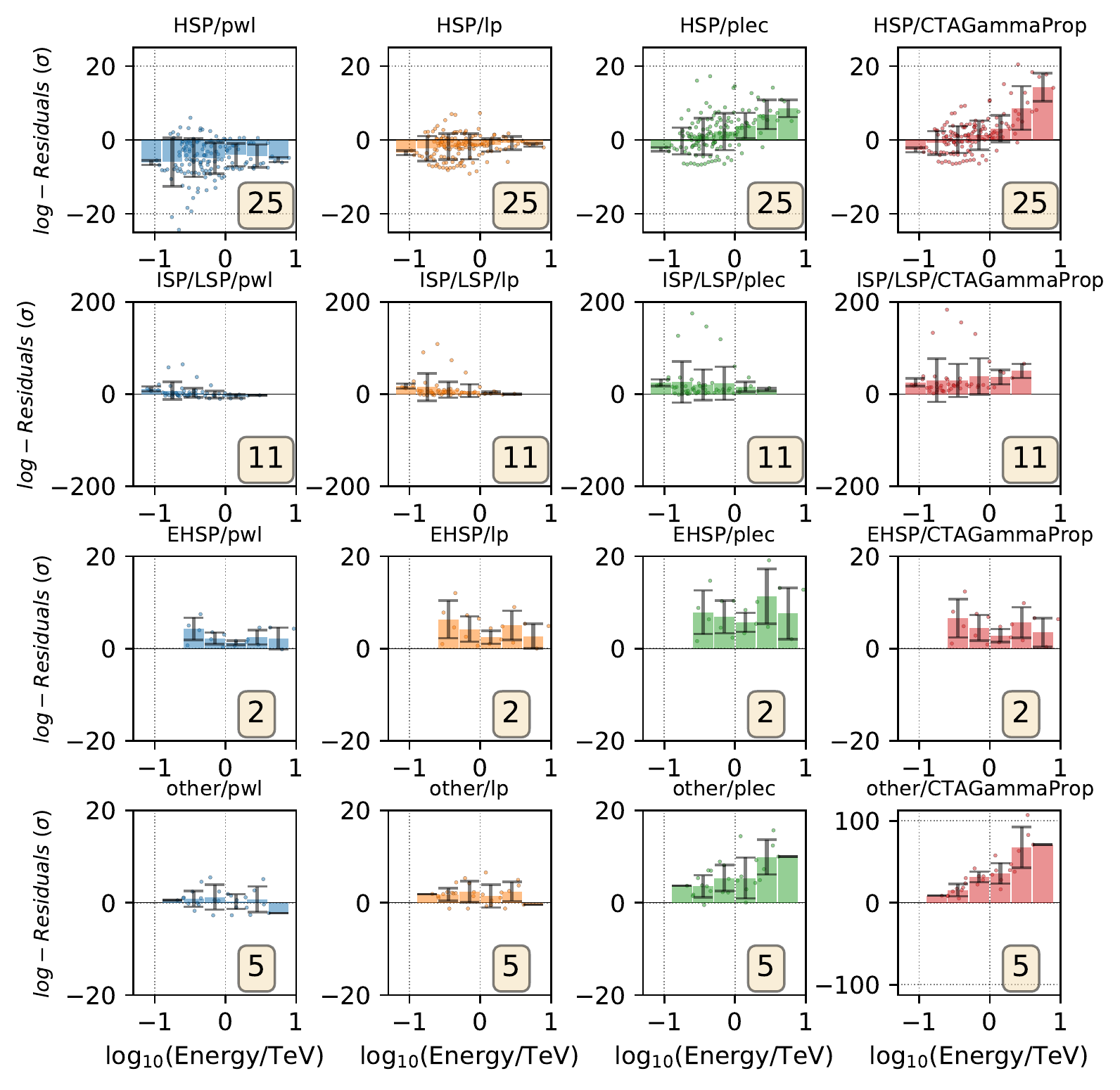}
    \caption{Residuals, calculated over the SED data in log scale. Each spectrum is fit with four different models (Power Law or PWL, Log Parabola or LP, Power Law with (sub-)exponential cut-off or PLEC or CTAGammaProp's schema (proposed in \citep{2021JCAP...02..048A}) for the lowest recorded state for each source. Note that the last case, other/CTAGammaProp, has a different vertical scale.}
    \label{fig:log-residuals}
\end{figure}

The analysis of flux residuals for a large sample of sources provides valuable information about the goodness of a given extrapolation scheme in predicting very high energy emission. We compute the residuals over the fluxes in logarithmic scale instead of linear scale. The estimated  `log-residuals' are shown in Figure \ref{fig:log-residuals} for each spectral type (rows) and each extrapolation scheme (columns).

Inspecting Figure \ref{fig:log-residuals}, we can conclude that the average spectrum of HSPs is best reproduced by a Log Parabola function. A simple Power Law slightly overestimates the VHE fluxes, resulting in mostly negative residuals. On the other hand, both PLEC and CTAGammaProp models provide reasonable estimates until $\sim 1\,\mathrm{TeV}$ but they tend to underestimate the fluxes above that energy. 

For ISP/LSPs, Figure \ref{fig:log-residuals} seems to indicate that the smallest and less biased residuals are obtained with either a PWL. However, the dispersion is very large (some points deviate more than $50\,\mathrm{\sigma}$), possibly due to the presence of FSRQs. These are very variable sources in $\gamma$-rays, being detected in many cases only when the source is in a high-state. This results in a large bias between the observations by {\it Fermi}-LAT, which consist on the average emission in several years, and the enhanced emission that is detected in just a few hours or days by VHE instruments during a flaring state. Finally, for EHSPs the small statistics that are available (only 2 sources in the catalogs considered) makes it impossible to draw any conclusions.

\section{Discussion}

This work is based on data extracted from catalogs. Its robustness depends on the completeness and accuracy of the results that are included in such catalogs. Within the HE $\gamma$-ray community there is a long tradition of releasing such catalogs, with {\it Fermi}-LAT being the leading figure in the field. This tradition does not exist in the Very High Energy community, which is divided into multiple collaborations, each having different ways of publicly releasing machine-readable scientific results. Publications are generally based on single sources, focusing on specific episodes or observing campaigns (e.g. flaring- or low-states, multi-instrument observations). 
Moreover, there have been attempts of making general catalogs out of published data (e.g.: TeVCat\footnote{\url{http://tevcat.uchicago.edu/}}, GammaCAT, VTSCat), but they have had generally limited success: TeVCat gathers a complete list of the detected sources in the VHE regime together with references to the relevant publications, but does not a machine-readable way of accessing published datasets. The collaborative initiative of GammaCAT seems like the right approach given its open nature, but is currently far from complete. VTSCat may be the most complete, fully-validated, and up-to-date release of public VHE datasets from an IACT collaboration, so perhaps it could mark the first step towards similar efforts performed by the rest of the collaborations in the field.

We are confident that the methods developed in this work can serve for future studies: they may help the current generation of IACTs to target particular extragalactic VHE emitters 
or to explore the population of sources potentially detectable by future VHE instruments. Even if for the current dataset we could only make reasonable predictions for HSPs, results for ISP/LPS or EHSPs might be drawn easily if more data and more complete samples were to be included.

\section{Conclusions}

We have developed a tool, {\tt ext2TeV}, to compare several extrapolation models coming from HE data with publicly available VHE spectra of AGNs. It currently implements catalog reading, source matching between catalogs, spectra extraction, estimation of the lowest state among the spectra available in VHE, calculation of residuals for VHE datasets based on models included in HE catalogs and EBL absorption. We performed a first scan based on the 4FGL, 4LAC, GammaCAT and VTSCat catalogs, compiling spectra for 43 blazars, most of them being HSP blazars. Out of the analysis of the residuals, we concluded that HSPs are best reproduced with a Log Parabola spectral shape, closely followed by {\it Fermi}-LAT's PLEC.

%
%
%

\end{document}